\documentclass{elsarticle}
\usepackage{amsmath,multirow}

\DeclareMathOperator{\Tr}{Tr}

\begin{document}

\begin{frontmatter}
\title{Correlators of heavy--light quark currents in HQET:\\
OPE at three loops%
\makebox(0,0){\raisebox{5cm}{\hspace{5.5cm}TTP21-054}}}
\author{K.\,G.~Chetyrkin}
\ead{konstantin.chetyrkin@partner.kit.edu}
\address{Institut f\"ur Theoretische Teilchenphysik,
Karlsruher Institut f\"ur Technologie, Karlsruhe, Germany}
\address{Institut f\"ur Theoretische Physik, Universit\"at
Regensburg, 93040~Regensburg, Germany,}
\address{II. Institut f\"ur Theoretische Physik,
Universit\"at  Hamburg, Luruper Chaussee~149, 22761~Hamburg, Germany}
\author{A.\,G.~Grozin}
\ead{A.G.Grozin@inp.nsk.su}
\address{Budker Institute of Nuclear Physics, Novosibirsk, Russia}
\address{Novosibirsk State University, Novosibirsk, Russia}
\begin{abstract}
Coefficient functions of the operator product expansion of
correlators of HQET heavy--light quark currents
are calculated up operators of dimension 4 up to to 3 loops.
\end{abstract}
\end{frontmatter}

\section{Heavy--light currents and their correlators in HQET}
\label{S:j}

QCD problems with a single heavy quark
$Q$ having momentum $P = M v + p$
($M$ is its on-shell mass, $v^2 = 1$)
can be described by heavy quark effective theory~\cite{Eichten:1989zv}
(HQET, see, e.\,g., \cite{Neubert:1993mb,Manohar:2000dt,Grozin:2004yc})
if its characteristic residual momentum is small ($p \ll M$),
and characteristic momenta of light quarks and gluons are also small.
QCD operators are expanded in $1/M$,
the coefficients are HQET operators of corresponding dimensionalities.
For example, QCD heavy--light quark currents at the leading order in $1/M$
are equal to the matching coefficients times the HQET heavy--light currents.
These matching coefficients are known at 2~\cite{Broadhurst:1994se,Grozin:1998kf}
and 3 loops~\cite{Bekavac:2009zc}.
Anomalous dimensions of all HQET heavy-light currents are the same
and known at 2~\cite{Ji:1991pr,Broadhurst:1991fz,Gimenez:1991bf}
and 3 loops~\cite{Chetyrkin:2003vi}.
Correlators of such currents at small distances can be calculated
using operator product expansion (OPE);
coefficient functions of operators up to dimension 3
are known up to 2 loops~\cite{Broadhurst:1991fc,Bagan:1991sg,Neubert:1991sp}.
The $\langle G^2\rangle$ contribution vanishes at 1 loop;
the $\langle G^3\rangle$ one is known at 1 loop~\cite{Broadhurst:1991fc}.
Contributions of quark condensates up to dimension 8
are known at tree level~\cite{Broadhurst:1991fc,Grozin:1994hd}.
Here we calculate the perturbative contribution expanded up to $m^4$
($m$ is the light-quark mass)
and condensate contributions up to dimension 4 at 3 loops.
The perturbative spectral densities of correlators of some QCD heavy--light currents
with $m=0$ in the threshold region at 3 loops were calculated ~\cite{Czarnecki:2001rh};
they are related to the HQET spectral density by the corresponding matching coefficients.

If our heavy quark is $b$, there are 2 different HQETs:
with $c$ quark and without it.
The heavy--light HQET currents in these 2 theories are related by the decoupling coefficient,
which is known up to 3 loops~\cite{Grozin:2006xm}.
The HQET current in HQET with $c$ is related to the QCD currents by the matching coefficients.
In this paper we shall work in HQET without $c$ quarks.
There are $n_l=3$ dynamic flavors ($u$, $d$, $s$)
and the static $b$ quark.

At the leading order in $1/M$ the heavy-quark
spin does not interact with gluon field.
We may rotate it at will without affecting physics
(heavy-quark spin symmetry~\cite{Isgur:1989ed}).
We may even switch it off (superflavor symmetry~\cite{Georgi:1990ak}).
We shall use the effective theory of a scalar static antiquark.
This particle has no antiparticle;
its field $\varphi^*$ contains only annihilation operators.
Its coordinate-space free propagator in the $v$ rest frame
is $\delta(\vec{x}) S_0(x^0)$ where $S_0(t) = -i \theta(t)$.
The momentum-space propagator $S_0(p) = 1/(p^0 + i0)$
does not depend on $\vec{p}$.
Static-quark lines cannot form loops.

We consider the current
\begin{equation}
j_0 = \varphi^*_0 q_0 = Z_j(\alpha_s(\mu)) j(\mu)\,.
\label{j:j}
\end{equation}
The correlator of 2 currents in the $v$ rest frame
\begin{equation}
\langle T j_0(x) \bar{\jmath}_0(0) \rangle = \delta(\vec{x}) \Pi_0(x^0)
\label{j:Pix}
\end{equation}
is non-zero only for $x^0 \ge 0$
(the symbol $T$ is superfluous: the product $\bar{\jmath}_0(0) j_0(x) = 0$).
The momentum-space correlator
\begin{equation}
\int d^d x\,\langle T j_0(x) \bar{\jmath}_0(0) \rangle e^{i p\cdot x} = \Pi_0(p^0)
\label{j:Pip}
\end{equation}
does not depend on $\vec{p}$.
They are related by the 1-dimensional Fourier transform
\begin{equation}
\Pi_0(\omega) = \int_0^\infty dt\,\Pi_0(t) e^{i \omega t}\,,\quad
\Pi_0(t) = \int_{-\infty}^{+\infty} \frac{d\omega}{2\pi} \Pi_0(\omega) e^{-i \omega t}\,.
\label{j:Fourier}
\end{equation}
The correlator $\Pi_0(\omega)$ has a cut from $0$ to $+\infty$,
the discontinuity gives the spectral density
\begin{equation}
\rho_0(\omega) = \frac{1}{2 \pi} \left[\Pi_0(\omega+i0) - \Pi_0(\omega-i0)\right]\,.
\label{j:rho}
\end{equation}
The correlator is expressed via the spectral density by the dispersion representation:
\begin{equation}
\Pi_0(\omega) = i \int_0^\infty \frac{d\nu\,\rho_0(\nu)}{\omega-\nu+i0}\,,\quad
\Pi_0(t) = \theta(t) \int_0^\infty d\omega\,\rho_0(\omega) e^{-i \omega t}\,.
\label{j:disp}
\end{equation}
We can analytically continue $\Pi_0(t)$ from $t>0$ to $t=-i\tau$, $\tau>0$
and obtain the Euclidean correlator
\begin{equation}
\Pi_0(\tau) = \int_0^\infty d\omega\,\rho_0(\omega) e^{- \omega \tau}\,.
\label{j:eucl}
\end{equation}
The spectral density can be reconstructed from it by the inverse Mellin transform
\begin{equation}
\rho_0(\omega) = \frac{1}{2 \pi i} \int_{a-i\infty}^{a+i\infty} d\tau\,\Pi_0(\tau) e^{\omega \tau}\,,
\label{j:mellin}
\end{equation}
where $a$ is to the right from all singularities of $\Pi_0(\tau)$.

Borel transform of the correlator $\Pi(\omega)$ is often used in sum rules.
In HQET it is defined by
\begin{equation}
\hat{B}_E F(\omega) = \lim_{k\to\infty} \frac{(-\omega)^{k+1}}{k!}
\biggl.\biggl(\frac{d}{d\omega}\biggr)^k F(\omega) \biggr|_{\omega = - E k}\,.
\label{j:Borel}
\end{equation}
It is equivalent to the correlator in imaginary time $\Pi(\tau)$.
For example, for the function
\begin{equation*}
F(\omega) = \frac{1}{(\nu - \omega - i0)^n}
\quad\text{it is}\quad
\hat{B}_E F(\omega) = \frac{e^{-\nu/E}}{\Gamma(n) E^{n-1}}\,.
\end{equation*}
The Fourier transform~(\ref{j:Fourier}) of $F(\omega)$ is
\begin{equation*}
F(t) = i \theta(t) \frac{(it)^{n-1}}{\Gamma(n)} e^{-i (\nu-i0) t}\,;
\end{equation*}
its analytical continuation from the half-axis $t>0$
to the half-axis $t = -i\tau$, $\tau > 0$ is
\begin{equation*}
F(\tau) = i \frac{\tau^{n-1}}{\Gamma(n)} e^{-\nu\tau}\,.
\end{equation*}
Therefore,
\begin{equation}
\hat{B}_E \Pi(\omega) = - i \Pi(\tau = 1/E)\,.
\label{j:Borel2}
\end{equation}

The static-antiquark propagator in a gluon field is
$\delta(\vec{x}_1-\vec{x}_0) S_0(x_1^0-x_0^0) \overline{[x_1,x_0]}$, where
\begin{equation}
\overline{[x_1,x_0]} = P \exp\left[- i g_0 \int_{x_0}^{x_1} d x_\mu A_0^\mu(x)\right]
\label{j:Wilson}
\end{equation}
is the Wilson line in the antiquark representation,
the integral is taken along the straight line from $x_0$ to $x_1$.
Therefore, the correlator can be written as
\begin{equation}
\Pi_0(t) = \langle q_0(vt) \overline{[vt,0]} \bar{q}_0(0) \rangle\,.
\label{j:WilsonPi}
\end{equation}
We can consider a more general object~\cite{Braun:2020ymy}
\begin{equation}
F_0(x) = \langle q_0(x) \overline{[x,0]} \bar{q}_0(0) \rangle\,,
\label{j:Pigenx}
\end{equation}
where $x$ is not necessarily timelike.
The bilocal vacuum average
\begin{equation}
\langle \bar{q}_0(0) [0,x] \Gamma q(x) \rangle = - \Tr \Gamma F_0(x)\,,
\label{j:biloc}
\end{equation}
where $[0,x]$ is the Wilson line in the quark (fundamental) representation,
and $\Gamma$ is a Dirac matrix.

The correlator of the $\overline{\text{MS}}$ renormalized currents $j(\mu)$
still contains ultraviolet (UV) divergences when $t=0$.
Subtracting these divergences
we obtain the renormalized correlator $\Pi(t;\mu)$.
The dispersion representation should contain 3 subtractions:
\begin{align}
&\Pi(\omega;\mu) = - i \omega^3 \int_0^\infty \frac{d\varepsilon\,\rho(\varepsilon;\mu)}{\varepsilon^3 (\varepsilon-\omega-i0)}
+ \sum_{n=0}^2 c_n \omega^n\,,
\nonumber\\
&\Pi(t;\mu) = \theta(t) \int_0^\infty d\omega\,\rho(\omega;\mu) e^{-i \omega t}
+ \sum_{n=0}^2 c_n i^n \delta^{(n)}(t)\,.
\label{j:disps}
\end{align}
Divergences of the correlator of renormalized currents
are in subtraction terms in~(\ref{j:disps}):
in coordinate space they are at $t=0$,
in momentum space they are polynomial in $\omega$.
More exactly, in dimensional regularization
only $c_2$ contains $1/\varepsilon^n$ divergences,
whereas power divergences in $c_{0,1}$ are not seen in this scheme.
The renormalized spectral density is simply given by
$\rho_0(\omega) = Z_j^2(\alpha_s(\mu)) \rho(\omega;\mu)$.

The correlator has 2 Dirac structures
\begin{equation}
\Pi = A + B \rlap/v\,.
\label{j:Dirac}
\end{equation}
It is convenient to introduce the currents with definite parities $P = \pm1$:
\begin{equation}
j_P = \frac{1 + P \rlap/v}{2} j\,.
\label{j:jP}
\end{equation}
Their correlators are
\begin{equation}
P \Pi_P \frac{1+P\rlap/v}{2}
\quad\text{where}\quad
\Pi_P = P(A + P B) = \frac{P}{4} \Tr (1 + P\rlap/v) \Pi
\label{j:PiP}
\end{equation}
(we shall see soon why it is convenient to introduce the factor $P$ here).

For sufficiently large $-\omega$ the operator product expansion (OPE) is valid
\begin{equation}
\Pi(\omega;\mu) = \sum_i C_i(\omega;\mu) \langle O_i(\mu) \rangle\,,
\label{j:OPE}
\end{equation}
where $O_i$ are all possible operators.
If the $q$ mass is small,
we can include operators with powers of $m(\mu)$ in the set $O_i$
and calculate the Wilson coefficients $C_i$ treating $q$ as massless.
Then the terms with even-dimensional $O_i$ have Dirac structure $\rlap/v$,
and those with odd-dimensional $O_i$ have the structure $1$.

Currently we are in the world where the antiquark $\bar{Q}$ has quantum numbers $0^+$.
Then $S$-wave $\bar{Q}q$ mesons have $j^P = \frac{1}{2}^+$,
and $P$-wave ones $\frac{1}{2}^-$ and $\frac{3}{2}^-$.
The currents $j_\pm$ have quantum numbers of $\frac{1}{2}^\pm$ mesons
(currents with quantum numbers of mesons with $j > \frac{1}{2}$ necessarily involve derivatives,
we don't consider them).
The matrix elements of our currents are
\begin{equation}
\langle 0 | j_P(\mu) | M \rangle = F(\mu) u\,,
\label{j:FP}
\end{equation}
where the meson states are normalized as
\begin{equation}
\langle M,\vec{p}^{\,\prime} | M,\vec{p} \rangle = (2\pi)^3 \delta(\vec{p}^{\,\prime} - \vec{p})
\label{j:nonrel}
\end{equation}
in the $v$ rest frame,
and $u$ is the Dirac wave function of the $\frac{1}{2}^P$ meson $M$
satisfying $\rlap/v u = P u$ and normalized as $u^+ u = 1$.
The contribution of the meson $M$ to the correlator $\Pi_P$ and its spectral density $\rho_P$ is
\begin{align}
&\Pi_M(t) = |F|^2 e^{- i \bar{\Lambda} t} \theta(t)\,,\quad
\Pi_M(\tau) = |F|^2 e^{- \bar{\Lambda} \tau}\,,
\nonumber\\
&\Pi_M(\omega) = \frac{i |F|^2}{\omega - \bar{\Lambda} + i0}\,,\quad
\rho_M(\omega) = |F|^2 \delta(\omega - \bar{\Lambda})\,,
\label{j:M}
\end{align}
where $\bar{\Lambda}$ is the residual energy of this meson.
If there are several mesons with given quantum numbers,
we get sums of contributions~(\ref{j:M});
sums become integrals in the continuum spectrum.

Now let's switch on the spin (and parity) $\frac{1}{2}^-$ of the static antiquark $\bar{Q}$
(still at $M = \infty$).
The static antiquark is now described by the field $\bar{h}$
satisfying $\bar{h} \rlap/v = - \bar{h}$.
The free propagator of this field contains the extra factor $(1 - \rlap/v)/2$
as compared to the scalar case.
The currents are
\begin{equation}
j_{\Gamma 0} = \bar{h}_0 \Gamma q_0\,;
\label{j:jGamma}
\end{equation}
in the $v$ rest frame the set of independent Dirac structures $\Gamma$ is
$1$, $\gamma^i$, $\gamma^{[i} \gamma^{j]}$, $\gamma^{[i} \gamma^{\vphantom{[]}j} \gamma^{k]}$,
where square brackets mean antisymmetrization.
Instead of this set, we can use
$1$, $\gamma_5^{\text{HV}}$, $\gamma^i$, $\gamma_5^{\text{HV}} \gamma^i$,
where $\gamma_5^{\text{HV}}$ is the 't~Hooft--Veltman $\gamma_5$.
In HQET renormalized currents with the anticommuting $\gamma_5^{\text{AC}}$
coincide with the corresponding currents with $\gamma_5^{\text{HV}}$,
because their anomalous dimensions are the same~\cite{Broadhurst:1994se}
(contrary to the QCD case).
Therefore, in the following we shall just use $\gamma_5$.
The correlator of $j_1^+$ and $j_2$ is
\begin{equation}
\Pi_{12} = - \Tr \bar{\Gamma}_1 \frac{1 - \gamma^0}{2} \Gamma_2 \Pi\,,
\label{j:Pi12}
\end{equation}
where $\Pi$ is the correlator with the scalar static antiquark,
and minus comes from the fermion loop.
The Dirac matrices $\Gamma$ either commute or anticommute with $\gamma^0$:
$\Gamma \gamma^0 = - P \gamma^0 \Gamma$.
Both $\Gamma_{1,2}$ must have the same $P$ (otherwise the correlator vanishes), and
\begin{equation}
\Pi_{12} = - \Pi_P \frac{1}{2} \Tr \bar{\Gamma}_1 \Gamma_2\,.
\label{j:Pi12a}
\end{equation}
The same formula works for the spectral densities.

\begin{table}[ht]
\begin{center}
\begin{tabular}{|cccc|}
\hline
$P$ & $\Gamma_1$ & $\Gamma_2$ & $\Pi_{12}$ \\
\hline
\multirow{2}{*}{$+1$} & $\gamma_5$ & $\gamma_5$ & $2 \Pi_+$ \\
& $\gamma^i$ & $\gamma^j$ & $2 \Pi_+ \delta^{ij}$ \\
\hline
\multirow{2}{*}{$-1$} & $1$ & $1$ & $2 \Pi_-$ \\
& $\gamma_5 \gamma^i$ & $\gamma_5 \gamma^j$ & $2 \Pi_- \delta^{ij}$ \\
\hline
\end{tabular}
\end{center}
\caption{Correlators of currents with spin $\frac{1}{2}$ heavy antiquark}
\label{T:c}
\end{table}

$S$-wave mesons with light-fields quantum numbers $j^P = \frac{1}{2}^+$ become degenerate doublets $0^-$, $1^-$;
$P$-wave ones with $j^P = \frac{1}{2}^-$, $\frac{3}{2}^-$ form degenerate doublets $0^+$, $1^+$ and $1^+$, $2^+$.
The currents with $\Gamma$ anticommuting with $\gamma^0$ ($\gamma_5$, $\gamma^i$: $P=+1$)
have quantum numbers of the $S$-wave $0^-$, $1^-$ mesons ($j^P = \frac{1}{2}^+$);
those with $\Gamma$ commuting with $\gamma^0$ ($1$, $\gamma_5 \gamma^i$: $P = -1$)
have quantum numbers of the $P$-wave $0^+$, $1^+$ mesons ($j^P = \frac{1}{2}^-$).
The spectral density of correlator of the currents with quantum numbers of $0^{\mp}$ mesons is $2 \rho_\pm$,
and for $1^\mp$ mesons it is $2 \rho_\pm \delta^{ij}$
(Table~\ref{T:c}, eq.~(\ref{j:Pi12a});
this is the reason why we introduced the factor $P$ in~(\ref{j:PiP})).
A $0^-$ meson contribution to the spectral density is $|F_{0^-}|^2 \delta(\omega-\bar{\Lambda})$;
for $1^-$ one it is $|F_{1^-}|^2 \delta^{ij} \delta(\omega-\bar{\Lambda})$, where
\begin{equation}
\langle 0 | \bar{h} \gamma_5 q | 0^- \rangle = F_{0^-}(\mu)\,,\quad
\langle 0 | \bar{h} \vec{\gamma} q | 1^- \rangle = F_{1^-}(\mu) \vec{e}
\label{j:F01}
\end{equation}
($\vec{e}$ is the polarization vector of the $1^-$ meson).
Therefore
\begin{equation}
F_{0^-}(\mu) = F_{1^-}(\mu) = \sqrt{2} F(\mu)
\label{j:spinsym}
\end{equation}
(this is an example of the heavy-quark spin symmetry).
The case of $0^+$, $1^+$ mesons is similar.

Usually the relativistic normalization of one-particle states is used:
\begin{equation}
{}_r\!\langle M, P' | M, P \rangle\!_r = (2\pi)^3 2P^0 \delta(\vec{P}' - \vec{P})
\label{j:rel}
\end{equation}
(it becomes meaningless when the meson mass $M \to \infty$, and thus is not usable in HQET;
in the meson rest frame $|M\rangle\!_r = \sqrt{2M} |M\rangle$).
The spin-symmetry result~(\ref{j:spinsym}) can be written in a completely Lorentz-invariant way:
\begin{align}
&\langle 0 | \bar{h} \Gamma q | M \rangle\!_r = \sqrt{M} F(\mu) \Tr \Gamma \mathcal{M}\,,\quad
\mathcal{M} = \frac{1 + \rlap/v}{2} \times
\left\{\begin{array}{ll}\gamma_5&\;\text{for }0^-\\\rlap/e&\;\text{for }1^-\end{array}\right.\,,
\nonumber\\
&\rlap/v \mathcal{M} = - \mathcal{M} \rlap/v = \mathcal{M}\,.
\label{j:spinsym2}
\end{align}
For example ($P^\mu = M v^\mu$ is the meson momentum)
\begin{align*}
&\langle 0 | \bar{h} \gamma_5 \gamma^\mu q | 0^- \rangle\!_r = \frac{2 F(\mu)}{\sqrt{M}} P^\mu\,,\quad
\langle 0 | \bar{h} \gamma_5 q | 0^- \rangle\!_r = \frac{2 F(\mu)}{\sqrt{M}} M\,,\\
&\langle 0 | \bar{h} \gamma^\mu q | 1^- \rangle\!_r = \frac{2 F(\mu)}{\sqrt{M}} M e^\mu\,,\quad
\langle 0 | \bar{h} \tfrac{1}{2}[\gamma^\mu,\gamma^\nu] q | 1^- \rangle\!_r = \frac{2 F(\mu)}{\sqrt{M}} (e^\mu P^\nu - e^\nu P^\mu)\,.
\end{align*}
Similarly, for $0^+$, $1^+$ mesons ($j^P = \frac{1}{2}^-$)
\begin{equation}
\mathcal{M} = \frac{1 - \rlap/v}{2} \times
\left\{\begin{array}{ll}1&\;\text{for }0^+\\\gamma_5 \rlap/e&\;\text{for }1^+\end{array}\right.\,,\quad
\rlap/v \mathcal{M} = \mathcal{M} \rlap/v = - \mathcal{M}\,.
\label{j:spinsym3}
\end{equation}
Of course, phases of $|M\rangle$ states (and hence of $\mathcal{M}$) can be redefined.

The vacuum average~(\ref{j:Pigenx}) for a general (timelike or spacelike) $x$
has 2 Dirac structures
\begin{align}
&F_0(x) = - \frac{1}{4} \left[F_S(x^2) - i \rlap/x F_V(x^2)\right]\,,
\nonumber\\
&F_S(x^2) = \langle \bar{q}(0) [0,x] q(x) \rangle\,,\quad
F_V(x^2) = \frac{i}{x^2} \langle \bar{q}(0) [0,x] \rlap/x q(x) \rangle\,.
\label{j:genx}
\end{align}
In HQET $x = vt$, and the scalar functions $F_{S,V}$ have positive argument $x^2 = t^2$.
If $x$ is spacelike, the argument $x^2$ is negative.
When we analytically continue HQET results to $t = -i\tau$,
we obtain $F_{S,V}$ of negative argument $-\tau^2$.

\section{Perturbative contribution}
\label{S:p}

\begin{figure}[ht]
\begin{center}
\begin{picture}(69,33)
\put(16,26){\makebox(0,0){\includegraphics{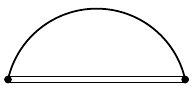}}}
\put(53,26){\makebox(0,0){\includegraphics{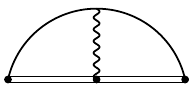}}}
\put(16,7){\makebox(0,0){\includegraphics{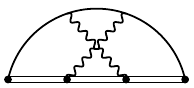}}}
\put(53,7){\makebox(0,0){\includegraphics{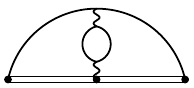}}}
\end{picture}
\end{center}
\caption{Perturbative contributions to the correlator.}
\label{F:p}
\end{figure}

Perturbative contributions to the correlator are shown in Fig.~\ref{F:p}: the
one-loop diagram; one of three two-loop ones; and two examples of three-loop
diagrams.
  
We use integration by parts (IBP) to reduce three-loop diagrams to master
integrals with the C++ program\footnote{We have also used the Mathematica
  program LiteRed~1.4 \cite{Lee:2012cn,Lee:2013mka} and the REDUCE package
  Grinder \cite{Grozin:2000jv} for testing purposes and the identification of
  the master integrals.}  FIRE6 \cite{Smirnov:2019qkx}. Generation of Feynman
diagram was done with QGRAF \cite{QGRAF} and evaluation of color factors with
the FORM \cite{Vermaseren:2000nd} package COLOR \cite{COLOR}.

There are three non-trivial master integrals.
Two of them are known exactly as hypergeometric functions with $\varepsilon$:
\cite{Beneke:1994sw} and~\cite{Grozin:2000jv,Grozin:2003ak};
for the last one, only a few terms of the $\varepsilon$ expansion are known~\cite{Czarnecki:2001rh},
but this is sufficient for our purpose.
This IBP procedure and the master integrals are reviewed in ~\cite{Grozin:2008tp}.

The renormalized perturbative correlator is
\begin{align}
&\Pi_P(\tau;\mu) = \frac{N_c}{\pi^2} \biggl\{
\frac{1}{2 \tau^3} \biggl[1
+ C_F \frac{\alpha_s}{4\pi} \biggl[6 L_\tau + 4 \biggl(\frac{\pi^2}{3} + 2\biggr)\biggr]
\nonumber\\
&\quad{} + C_F \left(\frac{\alpha_s}{4\pi}\right)^2 \biggl\{
C_F \biggl[18 L_\tau^2 + \biggl(\frac{40}{3} \pi^2 + 43\biggr) L_\tau
  - 8 \zeta_3 + \frac{8}{45} \pi^4 + \frac{52}{3} \pi^2 + \frac{153}{8}\biggr]
\nonumber\displaybreak\\
&\qquad{} + C_A \biggl[22 L_\tau^2 + \biggl(\frac{76}{9} \pi^2 + 75\biggr) L_\tau
  - 104 \zeta_3 - \frac{8}{45} \pi^4 - \frac{5}{27} \pi^2 + \frac{6413}{72}\biggr]
\nonumber\\
&\qquad{} - T_F n_l \biggl[8 L_\tau^2 + 4 \biggl(\frac{8}{9} \pi^2 + 7\biggr) L_\tau
  - 32 \zeta_3 - \frac{16}{27} \pi^2 + \frac{589}{18}\biggr]
\biggr\}\biggr]
\nonumber\\
&{} + P \frac{m}{4 \tau^2} \biggl[1
+ C_F \frac{\alpha_s}{4\pi} \biggl[12 L_\tau + 4 \biggl(\frac{\pi^2}{3} + 3\biggr)\biggr]
\nonumber\\
&\quad{} + C_F \left(\frac{\alpha_s}{4\pi}\right)^2 \biggl\{
C_F \biggl[72 L_\tau^2 + 2 \biggl(\frac{32}{3} \pi^2 + 71\biggr) L_\tau
  - 20 \zeta_3 + \frac{8}{45} \pi^4 + \frac{52}{3} \pi^2 + \frac{233}{4}\biggr]
\nonumber\\
&\qquad{} + C_A \biggl[44 L_\tau^2 + \frac{2}{3} \biggl(\frac{38}{3} \pi^2 + 205\biggr) L_\tau
  - 116 \zeta_3 - \frac{8}{45} \pi^4 + \frac{31}{27} \pi^2 + \frac{4981}{36}\biggr]
\nonumber\\
&\qquad{} - T_F n_l \biggl[16 L_\tau^2 + \frac{8}{3} \biggl(\frac{4}{3} \pi^2 + 17\biggr) L_\tau
  - 32 \zeta_3 - \frac{16}{27} \pi^2 + \frac{401}{9}\biggr]
\biggr\}\biggr]
\nonumber\\
&{} - \frac{m^2}{8 \tau} \biggl[1
+ 6 C_F \frac{\alpha_s}{4\pi} (3 L_\tau + 1)
\nonumber\\
&\quad{} + C_F \left(\frac{\alpha_s}{4\pi}\right)^2 \biggl\{
C_F \biggl[162 L_\tau^2 + \biggl(\frac{16}{3} \pi^2 + 109\biggr) L_\tau
  + 40 \zeta_3 - \frac{32}{45} \pi^4 - 12 \pi^2 + \frac{507}{8}\biggr]
\nonumber\\
&\qquad{} + C_A \biggl[66 L_\tau^2 - \biggl(\frac{4}{3} \pi^2 - 125\biggr) L_\tau
  - 22 \zeta_3 - \frac{4}{15} \pi^4 - \pi^2 + \frac{2789}{24}\biggr]
\nonumber\\
&\qquad{} - T_F n_l \biggl[24 L_\tau^2 + 36 L_\tau + \frac{229}{6}\biggr]\biggr\}\biggr]
\nonumber\\
&{} - 2 \frac{\sum m_i^2}{\tau} C_F T_F \left(\frac{\alpha_s}{4\pi}\right)^2 \biggl(\frac{\pi^2}{3} - 2\biggr)
\nonumber\\
&{} + P \frac{m^3}{8} \biggl[L_\tau
+ C_F \frac{\alpha_s}{4\pi} \biggl[12 L_\tau^2 + 10 L_\tau - 2 \biggl(\frac{2}{3} \pi^2 - 3\biggr)\biggr]
\nonumber\\
&\quad{} + C_F \left(\frac{\alpha_s}{4\pi}\right)^2 \biggl\{
C_F \biggl[96 L_\tau^3 + 2 \biggl(\frac{8}{3} \pi^2 + 59\biggr) L_\tau^2
  + \biggl(16 \zeta_3 - \frac{76}{3} \pi^2 + \frac{249}{2}\biggr) L_\tau
\nonumber\\
&\qquad\quad{} - 140 \zeta_5 + \frac{64}{3} \pi^2 \zeta_3 + 134 \zeta_3
  - \frac{26}{15} \pi^4 - \frac{40}{3} \pi^2 + \frac{91}{24}\biggr]
\nonumber\\
&\qquad{} + C_A \biggl[\frac{88}{3} L_\tau^3 - \frac{2}{3} (2 \pi^2 - 161) L_\tau^2
  - \biggl(16 \zeta_3 + \frac{97}{9} \pi^2 - \frac{889}{6}\biggr) L_\tau
\nonumber\\
&\qquad\quad{} + 60 \zeta_5 - \frac{20}{3} \pi^2 \zeta_3 + 89 \zeta_3
  - \frac{\pi^4}{45} - \frac{277}{27} \pi^2 + \frac{4357}{72}\biggr]
\nonumber\\
&\qquad{} - T_F n_l \biggl[\frac{32}{3} L_\tau^3 + \frac{104}{3} L_\tau^2
  - \frac{2}{3} \biggl(\frac{16}{3} \pi^2 - 71\biggr) L_\tau
  + 32 \zeta_3 - \frac{80}{27} \pi^2 + \frac{245}{18}\biggr]\biggr\}\biggr]
\nonumber\\
&{} - P m \Bigl(\sum m_i^2\Bigr) C_F T_F \left(\frac{\alpha_s}{4\pi}\right)^2
\biggl(6 L_\tau + \frac{\pi^2}{3} - \frac{2}{3}\biggr)
\nonumber\\
&{} - \frac{m^4 \tau}{32} \biggl[L_\tau - \frac{1}{4}
+ C_F \frac{\alpha_s}{4\pi} \biggl[18 L_\tau^2 - \frac{17}{2} L_\tau - \frac{8}{3} \pi^2 + \frac{45}{2}\biggr]
\nonumber\\
&\quad{} + C_F \left(\frac{\alpha_s}{4\pi}\right)^2 \biggl\{
C_F \biggl[186 L_\tau^3 + \biggl(\frac{16}{3} \pi^2 - \frac{259}{2}\biggr) L_\tau^2
  + \biggl(16 \zeta_3 - \frac{212}{3} \pi^2 + \frac{4155}{8}\biggr) L_\tau
\nonumber\\
&\qquad\quad{} - 420 \zeta_5 + 64 \pi^2 \zeta_3 + 351 \zeta_3
  - \frac{20}{9} \pi^4 - \frac{113}{9} \pi^2 - \frac{10609}{32}\biggr]
\nonumber\displaybreak\\
&\qquad{} + C_A \biggl[\frac{154}{3} L_\tau^3 - \biggl(\frac{4}{3} \pi^2 - \frac{63}{2}\biggr) L_\tau^2
  - \biggl(16 \zeta_3 + \frac{200}{9} \pi^2 - \frac{2603}{8}\biggr) L_\tau
\nonumber\\
&\qquad\quad{} + 180 \zeta_5 - 20 \pi^2 \zeta_3 + 179 \zeta_3 - \frac{\pi^4}{15} + \frac{835}{108} \pi^2 - \frac{64801}{288}\biggr]
\nonumber\\
&\qquad{} - T_F n_l \biggl[\frac{56}{3} L^3 + 2 L^2 - \biggl(\frac{64}{9} \pi^2 - \frac{235}{2}\biggr) L
  + 64 \zeta_3 + \frac{32}{27} \pi^2 - \frac{6137}{72}\biggr]\biggr\}\biggr]
\nonumber\\
&{} + m^2 \frac{\sum m_i^2}{4} \tau C_F T_F \left(\frac{\alpha_s}{4\pi}\right)^2
\biggl(6 L_\tau + \frac{17}{6}\biggr)
\nonumber\\
&{} - \frac{\sum m_i^4}{4} \tau C_F T_F \left(\frac{\alpha_s}{4\pi}\right)^2
\biggl[\biggl(\frac{\pi^2}{3} - \frac{11}{4}\biggr) L_\tau - 3 \zeta_3 - \frac{\pi^2}{2} + \frac{65}{8}\biggr]
\nonumber\\
&{} + \mathcal{O}(m^5,\alpha_s^3) \biggr\}\,,
\label{p:Pi}
\end{align}
where $\alpha_s = \alpha_s^{(n_l)}(\mu)$,
$m = m^{(n_l)}(\mu)$ is the mass of the quark $q$ in our current~(\ref{j:j}),
$m_i = m_i^{(n_l)}(\mu)$ are all light-flavor masses (see the last diagram in Fig.~\ref{F:p}),
and
\begin{equation}
L_\tau = \log\frac{\mu \tau e^{\gamma_E}}{2}\,.
\label{p:Lt}
\end{equation}

The coefficient functions $C_{m^n}(\mu)$ with $n=0$, $1$, $2$
satisfy simple renormalization group (RG) equations
(while $m^3$ mixes with $\bar{q}q$,
and $m^4$ mixes with $m\bar{q}q$ and $G^2$, Sect.~\ref{S:qg}).
Its solution is
\begin{align}
&C_{m^n} \sim \exp\biggl\{
\frac{\alpha_s}{4\pi} \bigl(- 2 \gamma_0 L_\tau + c_1\bigr)
+ \left(\frac{\alpha_s}{4\pi}\right)^2
\bigl[- 2 \beta_0 \gamma_0 L_\tau^2 - 2 (\gamma_1 - \beta_0 c_1) L_\tau + c_2\bigr]
\nonumber\\
&{} + \mathcal{O}(\alpha_s^3)\biggr\}\,,\quad
\gamma_k = 2 \gamma_{jk} - n \gamma_{mk}\,,\quad
\gamma_0 = - 6 C_F (n+1)\,.
\label{p:RG}
\end{align}
Here
\begin{align*}
&\gamma_a(\alpha_s) = \frac{d\log Z_a}{d\log\mu} = \sum_{n=0}^\infty \gamma_{an} \left(\frac{\alpha_s}{4\pi}\right)^{n+1}\quad
\mbox{($a = j$, $m$)}\,,\\
&\beta(\alpha_s) = \frac{1}{2} \frac{d\log Z_\alpha}{d\log\mu} = \sum_{n=0}^\infty \beta_n \left(\frac{\alpha_s}{4\pi}\right)^{n+1}\,.
\end{align*}
Our results~(\ref{p:Pi}) for $n=0$, $1$, $2$ satisfy this condition.

The renormalized spectral density of the OPE terms having dimensionalities $\le2$ is
\begin{align}
&\rho^{d\le2}_P(\omega;\mu) = \frac{N_c}{\pi^2} \biggl\{
\frac{\omega^2}{4} \biggl[1
- C_F \frac{\alpha_s}{4\pi} \biggl(6 L_\omega - \frac{4}{3} \pi^2 - 17\biggr)
\nonumber\\
&\quad{} + C_F \left(\frac{\alpha_s}{4\pi}\right)^2 \biggl\{
C_F \biggl[18 L_\omega^2 - \biggl(\frac{40}{3} \pi^2 + 97\biggr) L_\omega
- 8 \zeta_3 + \frac{8}{45} \pi^4 + \frac{103}{3} \pi^2 + \frac{1173}{8}\biggr]
\nonumber\displaybreak\\
&\qquad{} + C_A \biggl[22 L_\omega^2 - \biggl(\frac{76}{9} \pi^2 + 141\biggr) L_\omega
- 104 \zeta_3 - \frac{8}{45} \pi^4 + \frac{238}{27} \pi^2 + \frac{20057}{72}\biggr]
\nonumber\\
&\qquad{} - T_F n_l \biggl[8 L_\omega^2 - 4 \biggl(\frac{8}{9} \pi^2 + 13\biggr) L_\omega
- 32 \zeta_3 + \frac{92}{27} \pi^2 + \frac{1849}{18}\biggr]\biggr\}\biggr]
\nonumber\\
&{} + P \frac{m \omega}{4} \biggl[1
- 4 C_F \frac{\alpha_s}{4\pi} \biggl(3 L_\omega - \frac{\pi^2}{3} - 6\biggr)
\nonumber\\
&\quad{} + 4 C_F \left(\frac{\alpha_s}{4\pi}\right)^2 \biggl\{
C_F \biggl[18 L_\omega^2 - \biggl(\frac{16}{3} \pi^2 + \frac{143}{2}\biggr) L_\omega
- 5 \zeta_3 + \frac{2}{45} \pi^4 + \frac{20}{3} \pi^2 + \frac{1377}{16}\biggr]
\nonumber\\
&\qquad{} + C_A \biggl[11 L_\omega^2 - \frac{1}{3} \biggl(\frac{19}{3} \pi^2 + \frac{337}{2}\biggr) L_\omega
- 29 \zeta_3 - \frac{2}{45} \pi^4 + \frac{61}{108} \pi^2 + \frac{13069}{144}\biggr]
\nonumber\\
&\qquad{} - T_F n_l \biggl[4 L_\omega^2 - \frac{2}{3} \biggl(\frac{4}{3} \pi^2 + 29\biggr) L_\omega
- 8 \zeta_3 + \frac{2}{27} \pi^2 + \frac{1097}{36}\biggr]\biggr\}\biggr]
\nonumber\\
&{} - \frac{m^2}{8} \biggl[1
- 6 C_F \frac{\alpha_s}{4\pi} (3 L_\omega - 1)
\nonumber\\
&\quad{} + 2 C_F \left(\frac{\alpha_s}{4\pi}\right)^2 \biggl\{
C_F \biggl[81 L_\omega^2 - \biggl(\frac{8}{3} \pi^2 + \frac{109}{2}\biggr) L_\omega
+ 20 \zeta_3 - \frac{16}{45} \pi^4 - \frac{39}{2} \pi^2 + \frac{507}{16}\biggr]
\nonumber\\
&\qquad{} + C_A \biggl[33 L_\omega^2 + \biggl(\frac{2}{3} \pi^2 - \frac{125}{2}\biggr) L_\omega
- 11 \zeta_3 - \frac{2}{15} \pi^4 - 6 \pi^2 + \frac{2789}{48}\biggr]
\nonumber\\
&\qquad{} - T_F n_l \biggl[12 L_\omega^2 - 18 L_\omega - 2 \pi^2 + \frac{229}{12}\biggr]\biggr\}\biggr]
\nonumber\\
&{} - \frac{2}{3} \bigl(\sum m_i^2\bigr) C_F T_F \left(\frac{\alpha_s}{4\pi}\right)^2 (\pi^2 - 6)
+ \mathcal{O}(\alpha_s^3) \biggr\}\,,
\label{p:rho}
\end{align}
where
\begin{equation}
L_\omega = \log\frac{2\omega}{\mu}\,.
\label{p:L}
\end{equation}
Terms up to two loops agree with~\cite{Broadhurst:1991fc};
the remaining ones are new.
Multiplying the leading $m^0$ term in the HQET spectral density~(\ref{p:rho})
by the corresponding matching coefficients~\cite{Broadhurst:1994se,Grozin:1998kf},
we reproduce the leading $\delta^0$ terms in the 3-loop QCD spectral densities~(10), (14)
in~\cite{Czarnecki:2001rh}.

\section{Quark and gluon condensates (dimensions 3 and 4)}
\label{S:qg}

\begin{figure}[ht]
\begin{center}
\begin{picture}(106,33)
\put(16,26){\makebox(0,0){\includegraphics{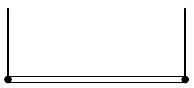}}}
\put(53,26){\makebox(0,0){\includegraphics{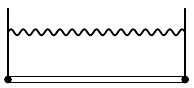}}}
\put(16,7){\makebox(0,0){\includegraphics{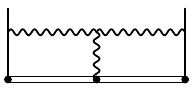}}}
\put(53,7){\makebox(0,0){\includegraphics{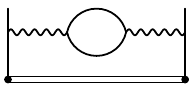}}}
\put(90,7){\makebox(0,0){\includegraphics{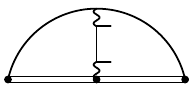}}}
\end{picture}
\end{center}
\caption{Quark-condensate contributions to the correlator.}
\label{F:Q}
\end{figure}

Some 0-, 1-, and 2-loop diagrams for the quark condensate contribution
are shown in Fig.~\ref{F:Q}.
Starting from 2 loops (the last diagram in the figure)
contributions proportional to the singlet sum $\sum m_i \langle \bar{q}_i q_i \rangle$ appear.
Our result for the coordinate-space correlator\footnote{
We have used  the well-known method of
projectors~\cite{Gorishnii:1983su,Gorishnii:1986gn} for computation
of various condensate contributions
(at 1 loop a similar method was used in~\cite{Broadhurst:1984rr,Broadhurst:1985js}).}
is
\begin{align}
&\Pi^q_P(\tau;\mu) = - P \frac{\langle\bar{q}q\rangle}{4} \biggl\{1 + 6 C_F \frac{\alpha_s}{4\pi}
\nonumber\displaybreak\\
&\quad{} + C_F \left(\frac{\alpha_s}{4\pi}\right)^2 \biggl\{
2 C_F \biggl[4 \biggl(\frac{2}{3} \pi^2 - 1\biggr) L_\tau - 16 \zeta_3 + \frac{10}{3} \pi^2 + 11\biggr]
\nonumber\\
&\qquad{} - C_A \biggl[4 \biggl(\frac{\pi^2}{3} - 7\biggr) L_\tau - 8 \zeta_3 + \pi^2 - \frac{149}{3}\biggr]
- 16 T_F n_l \biggl(L_\tau + \frac{4}{3}\biggr)
\biggr\}
\nonumber\\
&\quad{} + C_F \left(\frac{\alpha_s}{4\pi}\right)^3 \biggl\{
C_F^2 \biggl[4 \biggl(18 \zeta_3 + \frac{4}{9} \pi^4 + \frac{8}{3} \pi^2 - 35\biggr) L_\tau
\nonumber\\
&\qquad\quad{} + \frac{1600}{3} \zeta_5 - \frac{928}{9} \pi^2 \zeta_3 - \frac{140}{3} \zeta_3
+ \frac{479}{135} \pi^4 - \frac{8}{9} \pi^2 + 157\biggr]
\nonumber\\
&\qquad{} + C_F C_A \biggl[\frac{176}{3} \biggl(\frac{2}{3} \pi^2 - 1\biggr) L_\tau^2
- 4 \biggl(141 \zeta_3 - \frac{4}{45} \pi^4 - \frac{902}{27} \pi^2 - \frac{737}{9}\biggr) L_\tau
\nonumber\\
&\qquad\quad{} - \frac{1}{3} \biggl(1216 \zeta_5 - \frac{424}{3} \pi^2 \zeta_3 + \frac{23654}{9} \zeta_3
- \frac{3799}{270} \pi^4 - \frac{27122}{81} \pi^2 - \frac{23669}{27}\biggr)\biggr]
\nonumber\\
&\qquad{} - C_A^2 \biggl[\frac{88}{3} \biggl(\frac{\pi^2}{3} - 7\biggr) L_\tau^2
- 4 \biggl(33 \zeta_3 + \frac{2}{15} \pi^4 - \frac{164}{27} \pi^2 + \frac{1409}{9}\biggr) L_\tau
\nonumber\\
&\qquad\quad{} - 72 \zeta_5 + 12 \pi^2 \zeta_3 - \frac{4856}{27} \zeta_3
+ \frac{199}{810} \pi^4 + \frac{4094}{243} \pi^2 - \frac{69583}{81}\biggr]
\nonumber\\
&\qquad{} - 2 C_F T_F n_l \biggl[\frac{32}{3} \biggl(\frac{2}{3} \pi^2 - 1\biggr) L_\tau^2
- 32 \biggl(3 \zeta_3 - \frac{22}{27} \pi^2 - \frac{25}{9}\biggr) L_\tau
\nonumber\\
&\qquad\quad{} - \frac{1}{27} \biggl(5936 \zeta_3 - \frac{100}{3} \pi^4 - \frac{3536}{9} \pi^2 - \frac{10253}{3}\biggr)\biggr]
\nonumber\\
&\qquad{} + 2 C_A T_F n_l \biggl[16 \biggl(\frac{\pi^2}{9} - 6\biggr) L_\tau^2
- 4 \biggl(6 \zeta_3 - \frac{32}{27} \pi^2 +  \frac{217}{3}\biggr) L_\tau
\nonumber\\
&\qquad\quad{} - \frac{1}{27} \biggl(2258 \zeta_3 - \frac{89}{15} \pi^4 - \frac{830}{9} \pi^2 + \frac{27736}{3}\biggr)\biggr]
\nonumber\\
&\qquad{} + \frac{32}{3} \bigl(T_F n_l\bigr)^2 \biggl(4 L_\tau^2 + \frac{32}{3} L_\tau + \frac{109}{9}\biggr)
\biggr\}\biggr\}
\nonumber\\
&{} + \frac{m \langle\bar{q}q\rangle \tau}{16} \biggl\{
1 + 2 C_F \frac{\alpha_s}{4\pi} \biggl(3 L_\tau - \frac{5}{2}\biggr)
\nonumber\displaybreak\\
&\quad{} + 4 C_F \left(\frac{\alpha_s}{4\pi}\right)^2 \biggl\{
C_F \biggl[\frac{9}{2} L_\tau^2 + \biggl(\frac{4}{3} \pi^2 - \frac{35}{4}\biggr) L_\tau
- 8 \zeta_3 + \frac{8}{3} \pi^2 - \frac{547}{32}\biggr]
\nonumber\\
&\qquad{}  + C_A \biggl[\frac{11}{2} L_\tau^2 - \frac{1}{3} \biggl(\pi^2 + \frac{61}{4}\biggr) L_\tau
+ 2 \zeta_3 - \frac{3}{4} \pi^2 + \frac{3415}{96}\biggr]
\nonumber\\
&\qquad{} - T_F n_l \biggl(2 L_\tau^2 - \frac{5}{3} L_\tau + \frac{335}{24}\biggr)
\biggr\}
\nonumber\\
&\quad{} + 4 C_F \left(\frac{\alpha_s}{4\pi}\right)^3 \biggl[
C_F^2 \biggl[9 L_\tau^3 + 2 (4 \pi^2 - 15) L_\tau^2
- \biggl(30 \zeta_3 - \frac{4}{9} \pi^4 - 4 \pi^2 + \frac{1393}{16}\biggr) L_\tau
\nonumber\\
&\qquad\quad{} + \frac{1}{3} \biggl(1450 \zeta_5 - \frac{712}{3} \pi^2 \zeta_3 + 145 \zeta_3
+ \frac{431}{180} \pi^4 + \frac{461}{6} \pi^2 - \frac{20141}{32}\biggr)\biggr]
\nonumber\\
&\qquad{} + C_F C_A \biggl[33 L_\tau^3 + \frac{1}{3} \biggl(\frac{70}{3} \pi^2 - \frac{403}{2}\biggr) L_\tau^2
\nonumber\\
&\qquad\quad{} - \biggl(129 \zeta_3 - \frac{4}{45} \pi^4 - \frac{2551}{54} \pi^2 + \frac{6997}{144}\biggr) L_\tau
\nonumber\\
&\qquad\quad{} - \frac{1}{3} \biggl(754 \zeta_5 - \frac{256}{3} \pi^2 \zeta_3 + \frac{10715}{9} \zeta_3
- \frac{8119}{1080} \pi^4 - \frac{16439}{324} \pi^2 - \frac{207275}{432}\biggr)\biggr]
\nonumber\\
&\qquad{} + C_A^2 \biggl[\frac{242}{9} L_\tau^3 - \frac{1}{9} \biggl(22 \pi^2 + \frac{365}{2}\biggr) L_\tau^2
+ \biggl(33 \zeta_3 + \frac{2}{15} \pi^4 - \frac{362}{27} \pi^2 + \frac{26281}{54}\biggr) L_\tau
\nonumber\\
&\qquad\quad{} + \frac{67}{4} \zeta_5 - \frac{5}{2} \pi^2 \zeta_3 + \frac{2951}{27} \zeta_3
- \frac{419}{1620} \pi^4 - \frac{10103}{3888} \pi^2 - \frac{675449}{7776}\biggr]
\nonumber\\
&\qquad{} - C_F T_F n_l \biggl[12 L_\tau^3 + \frac{4}{3} \biggl(\frac{8}{3} \pi^2 - 13\biggr) L_\tau^2
- \biggl(72 \zeta_3 - \frac{496}{27} \pi^2 - \frac{13}{36}\biggr) L_\tau
\nonumber\\
&\qquad\quad{} - \frac{1}{27} \biggl(3058 \zeta_3 - \frac{223}{15} \pi^4 - \frac{1264}{9} \pi^2 - \frac{42839}{24}\biggr)\biggr]
\nonumber\\
&\qquad{} - C_A T_F n_l \biggl[\frac{176}{9} L_\tau^3 - \frac{2}{9} (4 \pi^2 + 71) L_\tau^2
+ 4 \biggl(9 \zeta_3 - \frac{34}{27} \pi^2 + \frac{2486}{27}\biggr) L_\tau
\nonumber\\
&\qquad\quad{} + \frac{1}{27} \biggl(436 \zeta_3 - \frac{53}{30} \pi^4 - \frac{353}{18} \pi^2 - \frac{26521}{36}\biggr)\biggr]
\nonumber\\
&\qquad{} + \frac{\bigl(T_F n_l\bigr)^2}{3}
\biggl(\frac{32}{3} L_\tau^3 - \frac{40}{3} L_\tau^2 + \frac{1940}{9} L_\tau + 8 \zeta_3 - \frac{6395}{162}\biggr)
\biggr]
\biggr\}
\nonumber\\
&{} + \Bigl(\sum m_i \langle \bar{q}_i q_i\rangle\Bigr) \frac{\tau}{4} C_F T_F \left(\frac{\alpha_s}{4\pi}\right)^2
\biggl\{
\frac{1}{2} \biggl(\frac{4}{3} \pi^2 - 11\biggr)
\nonumber\\
&\quad{} + \frac{\alpha_s}{4\pi} \biggl[
C_F \biggl[5 \biggl(\frac{4}{3} \pi^2 - 11\biggr) L_\tau + 12 \zeta_3 + \frac{8}{45} \pi^4 - \frac{44}{9} \pi^2 + \frac{76}{3}\biggr]
\nonumber\\
&\qquad{} + \frac{C_A}{3}
\biggl[\biggl(\frac{100}{3} \pi^2 - 311\biggr) L_\tau - 256 \zeta_3 - \frac{7}{15} \pi^4 - \frac{211}{9} \pi^2 + \frac{3559}{6}\biggr]
\nonumber\\
&\qquad{} - 2 T_F n_l
\biggl[\frac{4}{3} \biggl(\frac{4}{3} \pi^2 - 11\biggr) L_\tau - 16 \zeta_3 - \frac{44}{27} \pi^2 + \frac{313}{9}\biggr]
\biggr]
\biggr\}\,,
\label{q:q}
\end{align}
where $\langle\bar{q}q\rangle$ is renormalized at $\mu$.
The terms up to 2 loops in the dimension-3 contribution agree with~\cite{Broadhurst:1991fc}.

Finiteness of the renormalized coefficient function $C_{\bar{q}q}$
provides an independent confirmation of $2 \gamma_j - \gamma_{\bar{q}q}$
at 3 loops~\cite{Chetyrkin:2003vi}.
This anomalous dimension vanishes at 1 loop;
$\gamma_{\bar{q}q} = -\gamma_m$, hence $C_{\bar{q}q}$ has the structure~(\ref{p:RG})
with $n=-1$.
We need one more term:
\begin{align}
&C_{\bar{q}q} \sim \exp\biggl\{
c_1 \frac{\alpha_s}{4\pi}
+ \left(\frac{\alpha_s}{4\pi}\right)^2
\bigl[- 2 (\gamma_1 - \beta_0 c_1) L_\tau + c_2\bigr]
\label{q:RG}\\
&{} + \left(\frac{\alpha_s}{4\pi}\right)^3
\bigl[- 4 \beta_0 (\gamma_1 - \beta_0 c_1) L_\tau^2
- 2 (\gamma_2 - 2 \beta_0 c_2 - \beta_1 c_1) L_\tau + c_3\bigr]
+ \mathcal{O}(\alpha_s^4)
\biggr\}\,,
\nonumber
\end{align}
where $\gamma_k = 2 \gamma_{jk} + \gamma_{mk}$, $\gamma_0 = 0$.
Hence the $\alpha_s$ term contains no $L_\tau$,
the $\alpha_s^2$ one contains $L_\tau^1$, etc.

The dimension-3 operators $O_3 = (m^3,m\sum m_i^2,\bar{q}q)^T$
satisfy the renormalization group equation~%
\cite{Spiridonov:1988md,Chetyrkin:1994ex,Chetyrkin:2016ruf,Chetyrkin:2018avf}
\begin{align}
&\frac{d O_3}{d\log\mu} + \gamma_3 O_3 = 0\,,\quad
\gamma_3 = \left(\begin{array}{ccc}
3 \gamma_m & 0 & 0 \\
0 & 3 \gamma_m & 0 \\
\gamma & \gamma' & - \gamma_m
\end{array}\right)\,,
\label{q:RG3}\\
&\gamma = - \frac{N_c}{4 \pi^2} \biggl\{2 + 8 C_F \frac{\alpha_s}{4\pi}
\nonumber\\
&\quad{}
+ C_F \bigl[C_F (96 \zeta_3 - 131) - C_A (48 \zeta_3 - 109) - 20 T_F n_l\bigr] \left(\frac{\alpha_s}{4\pi}\right)^2
+ \mathcal{O}(\alpha_s^3)\biggr\}\,,
\nonumber\\
&\gamma' = 24 \frac{N_c}{\pi^2} C_F T_F \left(\frac{\alpha_s}{4\pi}\right)^2 + \mathcal{O}(\alpha_s^3)\,.
\nonumber
\end{align}
Therefore, the coefficient functions $C_3 = (C_{m^3},C_{m\sum m_i^2},C_{\bar{q}q})^T$
satisfy the renormalization group equation
\begin{equation}
\frac{d C_3}{d\log\mu}
= \frac{\partial C_3}{\partial L_\tau} - 2 \beta \frac{\partial C_3}{\partial\log\alpha_s}
= (\gamma_3^T - 2 \gamma_j) C_3\,.
\label{q:RG3C}
\end{equation}
The dimension-4 operators $m O_3$ satisfy the renormalization group equation
similar to~(\ref{q:RG3}) but with the anomalous dimension $\gamma_3 + \gamma_m$.
Hence we obtain
\begin{align*}
&\biggl[\frac{\partial}{\partial L_\tau} - 2 \beta \frac{\partial}{\partial\log\alpha_s}
+ 2 \gamma_j - 3 \gamma_m\biggr] C_{m^3} = \gamma C_{\bar{q}q}\,,\\
&\biggl[\frac{\partial}{\partial L_\tau} - 2 \beta \frac{\partial}{\partial\log\alpha_s}
+ 2 \gamma_j - 3 \gamma_m\biggr] C_{m\sum m_i^2} = \gamma' C_{\bar{q}q}\,,\\
&\biggl[\frac{\partial}{\partial L_\tau} - 2 \beta \frac{\partial}{\partial\log\alpha_s}
+ 2 \gamma_j - 4 \gamma_m\biggr] C_{m^4} = \gamma C_{m\bar{q}q}\,,\\
&\biggl[\frac{\partial}{\partial L_\tau} - 2 \beta \frac{\partial}{\partial\log\alpha_s}
+ 2 \gamma_j - 4 \gamma_m\biggr] C_{m^2\sum m_i^2} = \gamma' C_{m\bar{q}q}\,.
\end{align*}
Our results~(\ref{p:Pi}), (\ref{q:q}) satisfy these equations.

\begin{figure}[ht]
\begin{center}
\begin{picture}(36,22)
\put(18,12.5){\makebox(0,0){\includegraphics{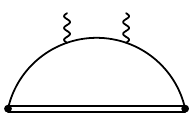}}}
\put(3,1){\makebox(0,0){$0$}}
\put(33,1){\makebox(0,0){$vt$}}
\end{picture}
\end{center}
\caption{One-loop gluon condensate contribution.}
\label{F:G}
\end{figure}

It is well known that the gluon condensate contribution vanishes at 1 loop.
In the fixed-point gauge the static quark does not interact with gluons,
and the only remaining diagram is shown in Fig.~\ref{F:G}.
But the $G^2$ correction to the massless quark propagator $S(x,0)$
vanishes after vacuum averaging~\cite{Novikov:1983gd}.
The 2- and 3-loop contributions are
\begin{align}
&\Pi^G_P(\tau;\mu) = \frac{\langle G^2\rangle \tau}{48} T_F \left(\frac{\alpha_s}{4\pi}\right)^2
\biggl\{
\biggl[C_F \biggl(\frac{4}{3} \pi^2 - 11\biggr) + \frac{C_A}{2} \biggl(\frac{4}{3} \pi^2 - 23\biggr)\biggr]
\nonumber\\
&\quad{} + 3 \frac{\alpha_s}{4\pi} \biggl\{
C_F^2 \biggl[2 \biggl(\frac{4}{3} \pi^2 - 11\biggr) L_\tau + 56 \zeta_3 + \frac{136}{135} \pi^4 - \frac{140}{9} \pi^2 - \frac{23}{2}\biggr]
\nonumber\\
&\qquad{} + \frac{C_F C_A}{9}
\biggl[\biggl(\frac{124}{3} \pi^2 - 449\biggr) L_\tau - 524 \zeta_3 - \frac{20}{3} \pi^4 + 100 \pi^2 + 167\biggr]
\nonumber\\
&\qquad{} + \frac{C_A^2}{9}
\biggl[11 \biggl(\frac{4}{3} \pi^2 - 23\biggr) L_\tau - 46 \zeta_3 + \frac{4}{5} \pi^4 + \frac{65}{6} \pi^2 - \frac{93}{2}\biggr]
\nonumber\\
&\qquad{} - \frac{2}{3} C_F T_F n_l
\biggl[\frac{4}{3} \biggl(\frac{4}{3} \pi^2 - 11\biggr) L_\tau - 16 \zeta_3 - \frac{44}{27} \pi^2 + \frac{259}{9}\biggr]
\nonumber\\
&\qquad{} - \frac{1}{3} C_A T_F n_l
\biggl[\frac{4}{3} \biggl(\frac{4}{3} \pi^2 - 23\biggr) L_\tau - 16 \zeta_3 - \frac{20}{27} \pi^2 + \frac{253}{9}\biggr]
\biggr\}
\biggr\}\,,
\label{g:g}
\end{align}
where $G^2=G^a_{\mu\nu} G^{a\mu\nu}$.
The anomalous dimension of this operator is~%
\cite{Collins:1976yq,Nielsen:1977sy,Spiridonov:1984br}
\begin{equation}
\gamma_{G^2} = - 2 \frac{d\beta}{d\log\alpha_s}\,,
\label{g:gamma}
\end{equation}
and hence the coefficient function must have the structure
\begin{equation}
C_{G^2} \sim \left(\frac{\alpha_s}{4\pi}\right)^2 \biggl\{1
+ \frac{\alpha_s}{4\pi} \bigl[2 (\beta_0 - \gamma_{j0}) L_\tau + c\bigr]
+ \mathcal{O}(\alpha_s^2)\biggr\}\,.
\label{g:RG}
\end{equation}
Our result~(\ref{g:g}) satisfies this condition.

The flavor-singlet dimension-4 operators $O_4=(\sum m_i^4,(\sum m_i^2)^2,\sum m_i \bar{q}_i q_i,\allowbreak G^2)^T$
satisfy the renormalization group equation~%
\cite{Spiridonov:1988md,Chetyrkin:1994ex,Chetyrkin:2016ruf,Chetyrkin:2018avf}
\begin{align}
&\frac{d O_4}{d\log\mu} + \gamma_4 O_4 = 0\,,
\label{g:RG4}\displaybreak\\
&\gamma_4 = \left(\begin{array}{cccc}
4 \gamma_m & 0 & 0 & 0 \\
0 & 4 \gamma_m & 0 & 0 \\
\gamma & \gamma' & 0 & 0 \\
- \frac{d\gamma}{d\log\alpha_s} & - \frac{d\gamma'}{d\log\alpha_s} & 4 \frac{d\gamma_m}{d\log\alpha_s} & - 2 \frac{d\beta}{d\log\alpha_s}
\end{array}\right)\,.
\nonumber
\end{align}
and the corresponding coefficient functions $C_4=(C_{\sum m_i^4},C_{(\sum m_i^2)^2},C_{\sum m_i \bar{q}_i q_i},\allowbreak C_{G^2})^T$
--- the equation
\begin{equation}
\frac{d C_4}{d\log\mu} = (\gamma_4^T - 2 \gamma_j) C_4\,;
\label{g:RG4C}
\end{equation}
hence,
\begin{align*}
&\biggl[\frac{\partial}{\partial L_\tau} - 2 \beta \frac{\partial}{\partial\log\alpha_s}
+ 2 \gamma_j - 4 \gamma_m\biggr] C_{\sum m_i^4}
= \gamma C_{\sum m_i \bar{q}_i q_i} - \frac{d\gamma}{d\log\alpha_s} C_{G^2}\,,\\
&\biggl[\frac{\partial}{\partial L_\tau} - 2 \beta \frac{\partial}{\partial\log\alpha_s}
+ 2 \gamma_j - 4 \gamma_m\biggr] C_{(\sum m_i^2)^2}
= \gamma' C_{\sum m_i \bar{q}_i q_i} - \frac{d\gamma'}{d\log\alpha_s} C_{G^2}\,,\\
&\biggl[\frac{\partial}{\partial L_\tau} - 2 \beta \frac{\partial}{\partial\log\alpha_s}
+ 2 \gamma_j\biggr] C_{\sum m_i \bar{q}_i q_i}
= 4 \frac{d\gamma_m}{d\log\alpha_s} C_{G^2}\,.
\end{align*}
The second equation here is satisfied trivially, because $C_{(\sum m_i^2)^2} = \mathcal{O}(\alpha_s^3)$.
Our results~(\ref{p:Pi}), (\ref{q:q}), (\ref{g:g}) satisfy these equations.

\section{Higher-dimensional condensates}
\label{S:h}

The tree diagram in Fig.~\ref{F:Q} can be written exactly in $x=vt$:
\begin{equation}
\Pi^q(t) = i \theta(t) \langle q(vt) \overline{[vt,0]} \bar{q}(0) \rangle\,.
\label{h:q}
\end{equation}
It is expressed via the bilocal quark condensate~\cite{Mikhailov:1986be}
which has 2 Dirac structures:
\begin{equation}
\langle q(x) \overline{[x,0]} \bar{q}(0) \rangle = - \frac{\langle\bar{q}q\rangle}{4}
\biggl[f_S(x^2) - \frac{i \rlap/x}{d} f_V(x^2)\biggr]\,.
\label{q:biloc}
\end{equation}
Its expansion in $x$ via local quark condensates is known
up to dimension 8~\cite{Grozin:1994hd}.
We use the bases of local condensates~\cite{Grozin:1986xh}
\begin{align}
&Q^3 = \langle \bar{q} q \rangle\,,\quad
Q^5 = i \langle \bar{q} G_{\mu\nu} \sigma^{\mu\nu} q \rangle\,,\quad
Q^6 = \langle \bar{q} \rlap/J q \rangle\,,
\nonumber\\
&Q^7_1 = \langle \bar{q} G_{\mu\nu} G^{\mu\nu} q \rangle\,,\quad
Q^7_2 = i \langle \bar{q} G_{\mu\nu} \tilde{G}^{\mu\nu} \gamma_5 q \rangle\,,
\nonumber\\
&Q^7_3 = \langle \bar{q} G_{\mu\lambda} G^\lambda{}_\nu \sigma^{\mu\nu} q \rangle\,,\quad
Q^7_4 = i \langle \bar{q} D_\mu J_\nu \sigma^{\mu\nu} q \rangle\,,
\nonumber\\
&A = i \langle \bar{q} D_\alpha D_\beta D_\gamma D_\delta D_\varepsilon
\gamma^{[\alpha} \gamma^{\vphantom{[}\beta} \gamma^{\vphantom{[}\gamma} \gamma^{\vphantom{[}\delta} \gamma^{\varepsilon]}
q \rangle\,,
\nonumber\\
&Q^8_1 = i \langle \bar{q} [[G_{\mu\lambda},G^\lambda{}_\nu]_+,D^\mu]_+ \gamma^\nu q \rangle\,,\quad
Q^8_2 = - \langle \bar{q} [[G_{\mu\lambda},\tilde{G}^\lambda{}_\nu],D^\mu]_+ \gamma^\nu \gamma_5 q \rangle\,,
\nonumber\\
&Q^8_3 = i \langle \bar{q} [\rlap{\,/}D G_{\mu\nu},G^{\mu\nu}] q \rangle\,,\quad
Q^8_4 = \langle \bar{q} D^2 \rlap/J q \rangle\,,
\nonumber\\
&Q^8_5 = i \langle \bar{q} [G_{\mu\nu},J^\mu] \gamma^\nu q \rangle\,,\quad
Q^8_6 = \langle \bar{q} [\tilde{G}_{\mu\nu},J^\mu]_+ \gamma^\nu \gamma_5 q \rangle\,,
\label{h:Q}
\end{align}
where $G_{\mu\nu} = g G^a_{\mu\nu} t^a$, $J_\mu = g J^a_\mu t^a$,
$J^a_\mu = D^\nu G^a_{\mu\nu} = g \sum \bar{q}_i \gamma_\mu t^a q_i$;
$\sigma_{\mu\nu} = \gamma_{[\mu} \gamma_{\nu]}$;
operators containing $\tilde{G}_{\mu\nu} = \frac{1}{2} \varepsilon_{\mu\nu\rho\sigma} G^{\rho\sigma}$
and $\gamma_5 = \frac{i}{4!} \varepsilon_{\alpha\beta\gamma\delta} \gamma^\alpha \gamma^\beta \gamma^\gamma \gamma^\delta$
are understood as short notations for the expressions from which both $\varepsilon$ tensors are eliminated using
$\varepsilon^{\mu\nu\rho\sigma} \varepsilon_{\alpha\beta\gamma\delta} = - 4!\,
\delta^\mu_{[\alpha} \delta^\nu_{\vphantom{[}\beta} \delta^\rho_{\vphantom{[}\gamma} \delta^\sigma_{\delta]}$.
The anomalous condensate $A$ does not vanish in the $\overline{\text{MS}}$ scheme;
it is a finite combination of dimension-8 gluon condensates~\cite{Grozin:1986xh}.

We obtain the contribution of bilinear quark condensates up to dimension 8 to the correlator at the tree level
\begin{align}
&\Pi^q_P(\tau) = - \frac{1}{4} \biggl\{ P Q^3 - \frac{\tau}{d} m Q^3
- P \frac{\tau^2}{2!\,d} \biggl[\frac{1}{2} Q^5 - m^2 Q^3\biggr]
\nonumber\\
&{} + \frac{\tau^3}{3!\,d(d+2)} \biggl[\frac{1}{2} Q^6 + \frac{3}{2} m Q^5 - 3 m^3 Q^3\biggr]
\nonumber\\
&{} + P \frac{\tau^4}{4!\,d(d+2)} \biggl[3 Q^7_1 - \frac{3}{2} Q^7_2 - 3 Q^7_3 + Q^7_4 - 2 m Q^6 - 3 m^2 Q^5 + 3 m^4 Q^3\biggr]
\nonumber\\
&{} - \frac{\tau^5}{5!\,d(d+2)(d+4)}
\biggl[5 A - \frac{5}{2} Q^8_2 + \frac{1}{4} Q^8_3 - \frac{1}{2} Q^8_4 - 3 Q^8_5 + 5 Q^8_6
\nonumber\\
&\quad{} + 5 m \bigl(3 Q^7_1 - Q^7_2 - 3 Q^7_3 + Q^7_4\bigr) - 15 m^2 \bigl(Q^6 + m Q^5 - m^3 Q^3\bigr)\biggr]
\nonumber\\
&{} + \mathcal{O}(\tau^6) \biggr\}\,.
\label{h:Pi}
\end{align}
The terms up to dimension 7 at $m=0$ agree with~\cite{Broadhurst:1991fc}.

\section{Conclusion}
\label{S:SR}

The results obtained here can be used for extracting numerical values of $F_P$
(and hence $f_B = f_{B^*}$, $f_{B_s}/f_B$ and similar quantities for $0^+$, $1^+$ mesons)
and $\bar{\Lambda}_P$
(and hence $m_{B_s}-m_B$, $m_{B(0^+)}-m_B$, $m_{B_s(0^+)}-m_{B(0^+)}$, $m_b$)
from HQET sum rules ($1/m_b$ corrections should be calculated separately).

For sufficiently small $\tau$ the correlator $\Pi_P(\tau;\mu)$ is given by the truncated OPE series
\begin{equation}
\Pi_P(\tau;\mu) = \int_0^\infty d\omega\,\rho_P^{d\le2}(\omega;\mu) e^{-\omega\tau} + \Pi_P^{d\ge3}(\tau;\mu)\,,
\label{SR:theor}
\end{equation}
where the coefficient functions are known as truncated series in $\alpha_s$.
On the other hand, we can represent it as
\begin{equation}
\Pi_P(\tau;\mu) = \int_0^\infty d\omega\,\rho_P(\omega;\mu) e^{-\omega\tau}\,,
\label{SR:phen}
\end{equation}
where the spectral density is given by the ground-state meson contribution~(\ref{j:M})
plus the continuum of excited states.
We can use the rough model of the continuum contribution~\cite{Shifman:1978bx}
\begin{equation}
\rho_P(\omega;\mu) = |F_P(\mu)|^2 \delta(\omega-\bar{\Lambda}_P)
+ \rho_P^{d\le2}(\omega;\mu) \theta(\omega-\omega_{cP})\,,
\label{SR:model}
\end{equation}
where $\omega_{cP}$ is the effective continuum threshold.
Equating these two expressions, we obtain the sum rule
\begin{equation}
|F_P(\mu)|^2 e^{-\bar{\Lambda}_P \tau}
= \int_0^{\omega_{cP}} d\omega\,\rho_P^{d\le2}(\omega;\mu) e^{-\omega\tau} + \Pi_P^{d\ge3}(\tau;\mu)\,.
\label{SR:SR}
\end{equation}
It is approximately valid at sufficiently large $\tau$,
where the continuum contribution is small,
and the uncertainty introduced by its rough model is not essential.
If there is a window of $\tau$ where both conditions are satisfied,
we can use this sum rule to extract an approximate value of $F_P(\mu)$.

Differentiating~(\ref{SR:SR}) in $\tau$ and dividing by~(\ref{SR:SR})
we obtain the sum rule for the ground-state residual energy
\begin{equation}
\bar{\Lambda}_P =
\frac{\int_0^{\omega_{cP}} d\omega\,\rho^{d\le2}_P(\omega;\mu) \omega e^{-\omega\tau} - d\Pi^{d\ge3}_P(\tau;\mu)/d\tau}%
{\int_0^{\omega_{cP}} d\omega\,\rho^{d\le2}_P(\omega;\mu) e^{-\omega\tau} + \Pi^{d\ge3}_P(\tau;\mu)}\,.
\label{SR:Lambda}
\end{equation}
The continuum thresholds $\omega_{cP}$ are tuned in such a way that the resulting $\bar{\Lambda}_P$
do not depend on $\tau$ in the region of applicability.

\section*{Acknowledgments}

We are grateful to S.\,V.~Mikhailov for useful comments.
The work of K.\,Ch.\ was supported in part by DFG grant CH~1479/2-1
and the  Research Unit FOR 2926, project number 40824754.
The work of A.\,G.\ has been supported by Russian Science Foundation under grant 20-12-00205.

\bibliography{corr}

\begin{thebibliography}{10}
\expandafter\ifx\csname url\endcsname\relax
  \def\url#1{\texttt{#1}}\fi
\expandafter\ifx\csname urlprefix\endcsname\relax\def\urlprefix{URL }\fi
\expandafter\ifx\csname href\endcsname\relax
  \def\href#1#2{#2} \def\path#1{#1}\fi

\bibitem{Eichten:1989zv}
E.~Eichten, B.~R. Hill, {An effective field theory for the calculation of
  matrix elements involving heavy quarks}, Phys. Lett. B 234 (1990) 511--516.
\newblock \href {https://doi.org/10.1016/0370-2693(90)92049-O}
  {\path{doi:10.1016/0370-2693(90)92049-O}}.

\bibitem{Neubert:1993mb}
M.~Neubert, {Heavy quark symmetry}, Phys. Rept. 245 (1994) 259--396.
\newblock \href {http://arxiv.org/abs/hep-ph/9306320}
  {\path{arXiv:hep-ph/9306320}}, \href
  {https://doi.org/10.1016/0370-1573(94)90091-4}
  {\path{doi:10.1016/0370-1573(94)90091-4}}.

\bibitem{Manohar:2000dt}
A.~V. Manohar, M.~B. Wise, {Heavy quark physics}, Vol.~10 of Camb. Monogr.
  Part. Phys. Nucl. Phys. Cosmol., {Cambridge university press}, Cambridge,
  2000.

\bibitem{Grozin:2004yc}
A.~G. Grozin, {Heavy quark effective theory}, Vol. 201 of Springer Tracts Mod.
  Phys., Springer, Berlin, 2004.
\newblock \href {https://doi.org/10.1007/b79301} {\path{doi:10.1007/b79301}}.

\bibitem{Broadhurst:1994se}
D.~J. Broadhurst, A.~G. Grozin, {Matching QCD and HQET heavy-light currents at
  two loops and beyond}, Phys. Rev. D 52 (1995) 4082--4098.
\newblock \href {http://arxiv.org/abs/hep-ph/9410240}
  {\path{arXiv:hep-ph/9410240}}, \href
  {https://doi.org/10.1103/PhysRevD.52.4082}
  {\path{doi:10.1103/PhysRevD.52.4082}}.

\bibitem{Grozin:1998kf}
A.~G. Grozin, {Decoupling of heavy quark loops in light-light and heavy-light
  quark currents}, Phys. Lett. B 445 (1998) 165--167.
\newblock \href {http://arxiv.org/abs/hep-ph/9810358}
  {\path{arXiv:hep-ph/9810358}}, \href
  {https://doi.org/10.1016/S0370-2693(98)01439-7}
  {\path{doi:10.1016/S0370-2693(98)01439-7}}.

\bibitem{Bekavac:2009zc}
S.~Bekavac, A.~G. Grozin, P.~Marquard, J.~H. Piclum, D.~Seidel, M.~Steinhauser,
  {Matching QCD and HQET heavy-light currents at three loops}, Nucl. Phys. B
  833 (2010) 46--63.
\newblock \href {http://arxiv.org/abs/0911.3356} {\path{arXiv:0911.3356}},
  \href {https://doi.org/10.1016/j.nuclphysb.2010.02.025}
  {\path{doi:10.1016/j.nuclphysb.2010.02.025}}.

\bibitem{Ji:1991pr}
X.-D. Ji, M.~Musolf, {Subleading logarithmic mass dependence in heavy meson
  form-factors}, Phys. Lett. B 257 (1991) 409--413.
\newblock \href {https://doi.org/10.1016/0370-2693(91)91916-J}
  {\path{doi:10.1016/0370-2693(91)91916-J}}.

\bibitem{Broadhurst:1991fz}
D.~J. Broadhurst, A.~G. Grozin, {Two loop renormalization of the effective
  field theory of a static quark}, Phys. Lett. B 267 (1991) 105--110.
\newblock \href {http://arxiv.org/abs/hep-ph/9908362}
  {\path{arXiv:hep-ph/9908362}}, \href
  {https://doi.org/10.1016/0370-2693(91)90532-U}
  {\path{doi:10.1016/0370-2693(91)90532-U}}.

\bibitem{Gimenez:1991bf}
V.~{Gim\'{e}nez}, {Two loop calculation of the anomalous dimension of the axial
  current with static heavy quarks}, Nucl. Phys. B 375 (1992) 582--622.
\newblock \href {https://doi.org/10.1016/0550-3213(92)90112-O}
  {\path{doi:10.1016/0550-3213(92)90112-O}}.

\bibitem{Chetyrkin:2003vi}
K.~G. Chetyrkin, A.~G. Grozin, {Three loop anomalous dimension of the
  heavy-light quark current in HQET}, Nucl. Phys. B 666 (2003) 289--302.
\newblock \href {http://arxiv.org/abs/hep-ph/0303113}
  {\path{arXiv:hep-ph/0303113}}, \href
  {https://doi.org/10.1016/S0550-3213(03)00490-5}
  {\path{doi:10.1016/S0550-3213(03)00490-5}}.

\bibitem{Broadhurst:1991fc}
D.~J. Broadhurst, A.~G. Grozin, {Operator product expansion in static quark
  effective field theory: Large perturbative correction}, Phys. Lett. B 274
  (1992) 421--427.
\newblock \href {http://arxiv.org/abs/hep-ph/9908363}
  {\path{arXiv:hep-ph/9908363}}, \href
  {https://doi.org/10.1016/0370-2693(92)92009-6}
  {\path{doi:10.1016/0370-2693(92)92009-6}}.

\bibitem{Bagan:1991sg}
E.~Bagan, P.~Ball, V.~M. Braun, H.~G. Dosch, {QCD sum rules in the effective
  heavy quark theory}, Phys. Lett. B 278 (1992) 457--464.
\newblock \href {https://doi.org/10.1016/0370-2693(92)90585-R}
  {\path{doi:10.1016/0370-2693(92)90585-R}}.

\bibitem{Neubert:1991sp}
M.~Neubert, {Heavy meson form-factors from QCD sum rules}, Phys. Rev. D 45
  (1992) 2451--2466.
\newblock \href {https://doi.org/10.1103/PhysRevD.45.2451}
  {\path{doi:10.1103/PhysRevD.45.2451}}.

\bibitem{Grozin:1994hd}
A.~G. Grozin, {Methods of calculation of higher power corrections in QCD}, Int.
  J. Mod. Phys. A 10 (1995) 3497--3529.
\newblock \href {http://arxiv.org/abs/hep-ph/9412238}
  {\path{arXiv:hep-ph/9412238}}, \href
  {https://doi.org/10.1142/S0217751X95001674}
  {\path{doi:10.1142/S0217751X95001674}}.

\bibitem{Czarnecki:2001rh}
A.~Czarnecki, K.~Melnikov, {Threshold expansion for heavy light systems and
  flavor off diagonal current current correlators}, Phys. Rev. D 66 (2002)
  011502.
\newblock \href {http://arxiv.org/abs/hep-ph/0110028}
  {\path{arXiv:hep-ph/0110028}}, \href
  {https://doi.org/10.1103/PhysRevD.66.011502}
  {\path{doi:10.1103/PhysRevD.66.011502}}.

\bibitem{Grozin:2006xm}
A.~G. Grozin, A.~V. Smirnov, V.~A. Smirnov, {Decoupling of heavy quarks in
  HQET}, JHEP 11 (2006) 022.
\newblock \href {http://arxiv.org/abs/hep-ph/0609280}
  {\path{arXiv:hep-ph/0609280}}, \href
  {https://doi.org/10.1088/1126-6708/2006/11/022}
  {\path{doi:10.1088/1126-6708/2006/11/022}}.

\bibitem{Isgur:1989ed}
N.~Isgur, M.~B. Wise, {Weak transition form-factors between heavy mesons},
  Phys. Lett. B 237 (1990) 527--530.
\newblock \href {https://doi.org/10.1016/0370-2693(90)91219-2}
  {\path{doi:10.1016/0370-2693(90)91219-2}}.

\bibitem{Georgi:1990ak}
H.~Georgi, M.~B. Wise, {Superflavor symmetry for heavy particles}, Phys. Lett.
  B 243 (1990) 279--283.
\newblock \href {https://doi.org/10.1016/0370-2693(90)90851-V}
  {\path{doi:10.1016/0370-2693(90)90851-V}}.

\bibitem{Braun:2020ymy}
V.~M. Braun, K.~G. Chetyrkin, B.~A. Kniehl, {Renormalization of parton
  quasi-distributions beyond the leading order: spacelike vs. timelike}, JHEP
  07 (2020) 161.
\newblock \href {http://arxiv.org/abs/2004.01043} {\path{arXiv:2004.01043}},
  \href {https://doi.org/10.1007/JHEP07(2020)161}
  {\path{doi:10.1007/JHEP07(2020)161}}.

\bibitem{Lee:2012cn}
R.~N. Lee, {Presenting LiteRed: a tool for the Loop InTEgrals REDuction} (12
  2012).
\newblock \href {http://arxiv.org/abs/1212.2685} {\path{arXiv:1212.2685}}.

\bibitem{Lee:2013mka}
R.~N. Lee, {LiteRed 1.4: a powerful tool for reduction of multiloop integrals},
  J. Phys. Conf. Ser. 523 (2014) 012059.
\newblock \href {http://arxiv.org/abs/1310.1145} {\path{arXiv:1310.1145}},
  \href {https://doi.org/10.1088/1742-6596/523/1/012059}
  {\path{doi:10.1088/1742-6596/523/1/012059}}.

\bibitem{Grozin:2000jv}
A.~G. Grozin, {Calculating three loop diagrams in heavy quark effective theory
  with integration by parts recurrence relations}, JHEP 03 (2000) 013.
\newblock \href {http://arxiv.org/abs/hep-ph/0002266}
  {\path{arXiv:hep-ph/0002266}}, \href
  {https://doi.org/10.1088/1126-6708/2000/03/013}
  {\path{doi:10.1088/1126-6708/2000/03/013}}.

\bibitem{Smirnov:2019qkx}
A.~V. Smirnov, F.~S. Chuharev, {FIRE6: Feynman Integral REduction with modular
  arithmetic} (2019).
\newblock \href {http://arxiv.org/abs/1901.07808} {\path{arXiv:1901.07808}}.

\bibitem{QGRAF}
P.~Nogueira, {Automatic Feynman graph generation}, J. Comput. Phys. 105 (1993)
  279--289.
\newblock \href {https://doi.org/10.1006/jcph.1993.1074}
  {\path{doi:10.1006/jcph.1993.1074}}.

\bibitem{Vermaseren:2000nd}
J.~A.~M. Vermaseren, {New features of FORM} (2000).
\newblock \href {http://arxiv.org/abs/math-ph/0010025}
  {\path{arXiv:math-ph/0010025}}.

\bibitem{COLOR}
T.~Van~Ritbergen, A.~N. Schellekens, J.~A.~M. Vermaseren, Group theory factors
  for feynman diagrams, Int. J. Mod. Phys. A 14~(1) (1999) 41--96.

\bibitem{Beneke:1994sw}
M.~Beneke, V.~M. Braun, {Heavy quark effective theory beyond perturbation
  theory: Renormalons, the pole mass and the residual mass term}, Nucl. Phys. B
  426 (1994) 301--343.
\newblock \href {http://arxiv.org/abs/hep-ph/9402364}
  {\path{arXiv:hep-ph/9402364}}, \href
  {https://doi.org/10.1016/0550-3213(94)90314-X}
  {\path{doi:10.1016/0550-3213(94)90314-X}}.

\bibitem{Grozin:2003ak}
A.~G. Grozin, {Lectures on multiloop calculations}, Int. J. Mod. Phys. A 19
  (2004) 473--520.
\newblock \href {http://arxiv.org/abs/hep-ph/0307297}
  {\path{arXiv:hep-ph/0307297}}, \href
  {https://doi.org/10.1142/S0217751X04016775}
  {\path{doi:10.1142/S0217751X04016775}}.

\bibitem{Grozin:2008tp}
A.~G. Grozin, {Higher radiative corrections in HQET}, in: {A. Ali and M.
  Ivanov} (Ed.), {Helmholz International Summer School on Heavy Quark Physics},
  Verlag Deutsches Elektronen-Synchrotron, 2008, pp. 55--88,
  {DESY-PROC-2009-07,
  http://www-library.desy.de/preparch/desy/proc/proc09-07.pdf}.
\newblock \href {http://arxiv.org/abs/0809.4540} {\path{arXiv:0809.4540}}.

\bibitem{Gorishnii:1983su}
S.~G. Gorishnii, S.~A. Larin, F.~V. Tkachov, {The algorithm for OPE coefficient
  functions in the MS scheme}, Phys. Lett. B 124 (1983) 217--220.
\newblock \href {https://doi.org/10.1016/0370-2693(83)91439-9}
  {\path{doi:10.1016/0370-2693(83)91439-9}}.

\bibitem{Gorishnii:1986gn}
S.~G. Gorishnii, S.~A. Larin, {Coefficient functions of asymptotic operator
  expansions in minimal subtraction scheme}, Nucl. Phys. B 283 (1987) 452.
\newblock \href {https://doi.org/10.1016/0550-3213(87)90283-5}
  {\path{doi:10.1016/0550-3213(87)90283-5}}.

\bibitem{Broadhurst:1984rr}
D.~J. Broadhurst, S.~C. Generalis, {Can mass singularities be minimally
  subtracted?}, Phys. Lett. B 142 (1984) 75--79.
\newblock \href {https://doi.org/10.1016/0370-2693(84)91139-0}
  {\path{doi:10.1016/0370-2693(84)91139-0}}.

\bibitem{Broadhurst:1985js}
D.~J. Broadhurst, S.~C. Generalis, {Dimension eight contributions to light
  quark QCD sum rules}, Phys. Lett. B 165 (1985) 175--180.
\newblock \href {https://doi.org/10.1016/0370-2693(85)90715-4}
  {\path{doi:10.1016/0370-2693(85)90715-4}}.

\bibitem{Spiridonov:1988md}
V.~P. Spiridonov, K.~G. Chetyrkin, {Nonleading mass corrections and
  renormalization of the operators $m\bar{\psi}\psi$ and $G^2_{\mu\nu}$}, Sov.
  J. Nucl. Phys. 47 (1988) 522--527, {Yad. Fiz. 47 (1988) 818--826}.

\bibitem{Chetyrkin:1994ex}
K.~G. Chetyrkin, J.~H. K{\"u}hn, {Quartic mass corrections to
  $R_{\text{had}}$}, Nucl. Phys. B 432 (1994) 337--350.
\newblock \href {http://arxiv.org/abs/hep-ph/9406299}
  {\path{arXiv:hep-ph/9406299}}, \href
  {https://doi.org/10.1016/0550-3213(94)90605-X}
  {\path{doi:10.1016/0550-3213(94)90605-X}}.

\bibitem{Chetyrkin:2016ruf}
K.~G. Chetyrkin, M.~F. Zoller, {Leading QCD-induced four-loop contributions to
  the \ensuremath{\beta}-function of the Higgs self-coupling in the SM and
  vacuum stability}, JHEP 06 (2016) 175.
\newblock \href {http://arxiv.org/abs/1604.00853} {\path{arXiv:1604.00853}},
  \href {https://doi.org/10.1007/JHEP06(2016)175}
  {\path{doi:10.1007/JHEP06(2016)175}}.

\bibitem{Chetyrkin:2018avf}
P.~A. Baikov, K.~G. Chetyrkin, {QCD vacuum energy in 5 loops}, PoS RADCOR2017
  (2018) 025.
\newblock \href {https://doi.org/10.22323/1.290.0025}
  {\path{doi:10.22323/1.290.0025}}.

\bibitem{Novikov:1983gd}
V.~A. Novikov, M.~A. Shifman, A.~I. Vainshtein, V.~I. Zakharov, {Calculations
  in external fields in quantum chromodynamics. Technical review}, Fortsch.
  Phys. 32 (1984) 585.

\bibitem{Collins:1976yq}
J.~C. Collins, A.~Duncan, S.~D. Joglekar, {Trace and dilatation anomalies in
  gauge theories}, Phys. Rev. D 16 (1977) 438--449.
\newblock \href {https://doi.org/10.1103/PhysRevD.16.438}
  {\path{doi:10.1103/PhysRevD.16.438}}.

\bibitem{Nielsen:1977sy}
N.~K. Nielsen, {The energy momentum tensor in a nonabelian quark gluon theory},
  Nucl. Phys. B 120 (1977) 212--220.
\newblock \href {https://doi.org/10.1016/0550-3213(77)90040-2}
  {\path{doi:10.1016/0550-3213(77)90040-2}}.

\bibitem{Spiridonov:1984br}
V.~P. Spiridonov, {Anomalous dimension of $G^2_{\mu\nu}$ and $\beta$ function},
  Tech. Rep. P-0378, IYaI,
  https://lib-extopc.kek.jp/preprints/PDF/1986/8601/8601315.pdf (1984).

\bibitem{Mikhailov:1986be}
S.~V. Mikhailov, A.~V. Radyushkin, {Nonlocal condensates and QCD sum rules for
  pion wave function}, JETP Lett. 43 (1986) 712, {Pisma Zh. Eksp. Teor. Fiz. 43
  (1986) 551}.

\bibitem{Grozin:1986xh}
A.~G. Grozin, Y.~F. Pinelis, {Contribution of higher gluon condensates to the
  light quark vacuum polarization}, Z. Phys. C 33 (1987) 419.
\newblock \href {https://doi.org/10.1007/BF01552548}
  {\path{doi:10.1007/BF01552548}}.

\bibitem{Shifman:1978bx}
M.~A. Shifman, A.~I. Vainshtein, V.~I. Zakharov, {QCD and Resonance Physics.
  Theoretical Foundations}, Nucl. Phys. B 147 (1979) 385--447.
\newblock \href {https://doi.org/10.1016/0550-3213(79)90022-1}
  {\path{doi:10.1016/0550-3213(79)90022-1}}.

\end{thebibliography}
\end{document}